\documentclass[sigconf]{acmart}

\usepackage{graphicx}
\usepackage{xurl}

\usepackage{amsmath,amssymb}

\usepackage{textcomp}
\usepackage{xcolor}
\usepackage{verbatim}
\usepackage{booktabs}
\usepackage{multirow}
\usepackage{multicol}
\usepackage[most]{tcolorbox}
\usepackage{colortbl}
\usepackage{microtype}
\usepackage{array}
\usepackage{listings}
\acmConference[SVM'26]
{Proceedings of the Fourth International Workshop on Software Vulnerability Management (SVM 2026), Rio De Janeiro,Brazil}
{April 12--18, 2026}
{Porto de Galinhas, Brazil}

\begin{document}
\title{LLMs in Code Vulnerability Analysis: A Proof of Concept}
\author{Shaznin Sultana}
    \affiliation{%
        \institution{Ohio University}
        \city{Athens}
        \state{Ohio}
        \country{USA}
}
\email{ss264525@ohio.edu}

\author{Sadia Afreen}
    \affiliation{%
        \institution{University of Cincinnati}
        \city{Cincinnati}
        \state{Ohio}
        \country{USA}
}
\email{afreensa@mail.uc.edu}

\author{Nasir U. Eisty}
\affiliation{%
	   \institution{University of Tennessee}
	   \city{Knoxville}
	   \state{TN}
	   \country{USA}
}
\email{neisty@utk.edu}
\date{October 2025}

\begin{abstract}
\textit{\textbf{Context:}} Traditional software security analysis methods struggle to keep pace with the scale and complexity of modern codebases, requiring intelligent automation to detect, assess, and remediate vulnerabilities more efficiently and accurately.
\textit{\textbf{Objective}:} This paper explores the incorporation of code-specific and general-purpose Large Language Models (LLMs) to automate critical software security tasks, such as identifying vulnerabilities, predicting the severity and access complexity rate, and generating fixes as a proof of concept. 
\textit{\textbf{Method}}: We evaluate five pairs of the most recent LLMs, which contain both code-based and general-purpose open-source models, on two recognized C/C++ vulnerability datasets, namely Big-Vul and Vul-Repair. 
Additionally, we show comparisons between fine-tuning and prompt-based approaches.
\textit{\textbf{Results}}: The results show that fine‑tuning uniformly outperforms both zero-shot and few‑shot approaches across all tasks and models. 
Notably, code‑specialized models excel in zero‑shot and few‑shot settings on complex tasks, but general‑purpose models remain nearly as effective. Discrepancies among CodeBLEU, CodeBERTScore, BLEU, and ChrF highlight the inadequacy of the current metrics for measuring repair quality.
\textit{\textbf{Conclusions}}: Through this study, we contribute to the resilience of the current software security community by investigating the potential of advanced LLMs. 

\end{abstract}


\maketitle

\section{Introduction}\label{sec:introduction}

Traditional vulnerability assessment tools, such as static analyzers, have limitations in coverage, programming language support, and vulnerability identification accuracy ~\cite{ghaleb_how_2020}. Some major problems developers encounter include the likelihood of bugs, the need for frequent configuration, and the difficulty of using tools ~\cite{beller_analyzing_2016, vassallo_how_2020}. It is challenging to detect complicated vulnerabilities manually or using traditional tools in large, correlated code bases. Modern software security requires intelligence that adapts rather than static tools and manual code reviews. Using automated vulnerability detection systems based on machine learning and deep learning to monitor, evaluate, and respond to potential threats is much more efficient than human methods~\cite{vuldetect_syslit,steenhoek_empirical_2023, adriana2024toward}.

With the tremendous development of LLMs~\cite{LLM_godfatherpaper} in large-scale text and code analysis, they are being investigated beyond general natural language processing (NLP) tasks in diverse areas~\cite{survey_llm, LLM_taxonomy}. 
LLMs are making great progress in software engineering~\cite {llm_in_SE}. 
It has led to further exploration about whether LLMs can complement critical fields like vulnerability analysis~\cite{repairComprehensice} or whether human experts are still mandatory to some extent~\cite{LLM_notyet}. 
However, LLMs' practical value, adaptability, and cost efficiency require a more comprehensive study beyond detecting vulnerabilities, such as analyzing their severity and impact. 

As a proof of concept for the SVM workflow, this study assesses open-source LLMs to streamline software security. While existing studies largely focus on vulnerability detection and repair, we introduce a more granular evaluation that benchmarks five pairs of code-specific versus general-purpose LLMs across four-stage pipeline. Starting with detection to risk assessment, where severity and access complexity predictions inform the prioritization of critical vulnerabilities, facilitating the repair generation. 
We focus on clarifying the benefits and limits of LLMs by comparing their performance to recognized metrics. In this process, we utilize two widely used large datasets, Big-vul~\cite{bigvul} and Vul-Repair~\cite{Chen_2023vulrepair}, on C/C++ programs to make the comparison. 
For the LLM selection, we chose five pairs of code-based and general open-source models. 

Through our experiments, we address the following research questions to uncover significant insights into the effectiveness of LLMs in contributing to code vulnerability analysis.

\begin{description}
    \item[\textbf{RQ1:} ] How does the effectiveness of fine-tuning compare to prompt-based approaches like zero-shot and few-shot learning?
    \item[\textbf{RQ2:} ] Do code-specific models consistently outperform general-purpose models across different approaches, or is it the other way around?
    \item[\textbf{RQ3:} ] Which pair of models shows superior performance relative to other pairs?
    \item[\textbf{RQ4:} ] Are existing code similarity metrics adequate for evaluating the capabilities of LLMs in code fix generation tasks?
\end{description}

The contributions of our study are as follows:

\begin{itemize}
    \item We conduct a thorough evaluation to determine if code-specialized models offer significant advantages over general-purpose models in vulnerability analysis.
    \item We compare prompt-based methods with fine-tuned approaches to assess whether the efficiency of prompting compensates for the higher resource demands of fine-tuning in handling complex vulnerabilities.
    \item We evaluate open-source LLMs to explore their potential as viable alternatives to proprietary models.
    \item We have made our code and dataset available at - \textcolor{blue}{\url{https://figshare.com/s/a06ec09cd1bd98e6dd45}}

\end{itemize}


\section{Related Works}\label{sec:background}
Recent advances in LLMs have inspired research into software vulnerability detection~\cite{LLMagent_2025}. Early efforts modified BERT or combined sequence–graph embeddings~\cite{hou_large_2023}, while Zhou et al.~\cite{zhou2024emergingresults} used GPT-3.5/4 with in-context learning to identify bugs. Beyond detection~\cite{vuldetectbench2024,enhancingreverseengineering2024,closingthegap2025, islam2024enhancingsourcecodesecurity, Multitask2024}, studies examined LLMs’ abilities in vulnerability description~\cite{enhancingreverseengineering2024, Multitask2024, islam2024enhancingsourcecodesecurity}, localization~\cite{closingthegap2025, Multitask2024}, repair~\cite{islam2024enhancingsourcecodesecurity, chatgpt2023}, classification~\cite{vuldetectbench2024,enhancingreverseengineering2024,closingthegap2025, chatgpt2023}, and root-cause analysis~\cite{vuldetectbench2024,enhancingreverseengineering2024}.

Most studies highlight fine-tuning as effective. Liu et al.~\cite{vuldetectbench2024} showed that larger models perform better on simple tasks but struggle with root-cause analysis. Islam et al.~\cite{islam2024enhancingsourcecodesecurity} improved repair by 12\% using SecRepair and reinforcement learning. Zhou et al.~\cite{review2024} offered a roadmap for adapting and deploying LLMs in workflows.

Prompt design also matters. Liu et al.~\cite{liu2024prompting} found that combining code details with CoT prompts boosts GPT-3.5 turbo. Mathews et al.~\cite{mathews_llbezpeky_2024} used retrieval-augmented prompts, emphasizing the use of structured pipelines. Manuel et al.~\cite{enhancingreverseengineering2024} fine-tuned on decompiled code but noted dataset bias.

LLMs also support repair generation~\cite{wu2023neuralnetworks, wang2024aigeneratedcodereallysafe, repairComprehensice, david2024enhancedrepair, zhou_large_2024}. Huang et al.~\cite{repairempirical2023} improved accuracy by marking bug locations, though models falter on long sequences. Zhou et al.~\cite{zhou2024multillmcollaborationdatacentric} localized bugs with $<start><end>$ tokens. Boyang et al.~\cite{yang2024multiobjectivefinetuningenhancedprogram} introduced MOREPAIR, optimizing repair and explanation jointly.

Hybrid approaches like IRIS~\cite{li2024llmassistedstaticanalysisdetecting} combine LLMs with static analysis, detecting 55 of 120 vulnerabilities and finding six new ones. Zhou et al.~\cite{zhou2024comparisonstaticapplicationsecurity} and Steenhoek et al.~\cite{closingthegap2025} compared static analyzers and LLMs, finding higher precision but limited developer trust.

\textbf{All studies indicate that LLMs exhibit promising traits that encourage further extensive research in this domain.}



\section{Methodology}\label{sec:methodology}
\subsection{LLMs Selection}
We selected two groups of the latest released LLMs: \textbf{code-specific models} and \textbf{general-purpose models}. 
We examined three popular leaderboards for LLMs, such as Chatbot Arena LLM Leaderboard~\cite{lmsys2024}, Trustbit LLM Benchmark~\cite{trustbit2024}, and CanAiCode Leaderboard~\cite{canaicode2024}, and cross-checked to select these LLMs from the top 20 open-source LLMs. Additionally, we chose each code-based LLM and its general version in order to conduct a comparative analysis of their performance. Table~\ref{tab:model_details} summarizes the models and the versions used in our experiment.

\subsection{Tasks Selection}

We examined the models on three key task categories: \textbf{Detection}, \textbf{Prediction}, and \textbf{Generation}, specifically within software code vulnerability. We selected these tasks to assess the models' abilities to not only identify vulnerabilities but also to analyze and mitigate them effectively. 

\subsubsection{\textbf{Detection}}
The models analyze a given code snippet to determine whether it contains a security vulnerability and what the vulnerability type is.
For this task, the models must be familiar with various programming techniques and identify patterns associated with common security flaws, such as buffer overflows and SQL injection.

\subsubsection{\textbf{Severity Prediction}}
Upon detecting a vulnerability, the models classify its severity level as high, medium, or low.
Severity prediction relies on the Common Vulnerability Scoring System (CVSS)~\cite{severity} as a guiding metric. Using CVSS-labeled data, models learn patterns linking vulnerability descriptions to specific scores or levels.
According to Le et al.~\cite{severity}, CVSS scores range
from 0 to 10 and are grouped into levels or categories to define the urgency and the required remediation. The groups are: 
\begin{itemize}
    \item \textbf{Low Severity:} 0.0 - 3.9 (minimal exploitability).
    \item \textbf{Medium Severity:} 4.0 - 6.9 (requires action but not immediately).
    \item \textbf{High Severity:} 7.0 - 10.0 (serious threats with immediate action and broad consequences).
\end{itemize}

Based on the CVSS score, we manually categorized data into Low, Medium, and High severity ranks. We feed these severity levels along with the CVSS score to the models for learning the pattern. It involves understanding how vulnerabilities can be exploited and being alert to potential consequences.

\subsubsection{\textbf{Access Complexity Classification}}
The models assess and categorize the potential access capability of a detected vulnerability as high, medium, or low. 
For this task, we utilized the Access Complexity score~\cite{bigvul, CWE} to classify the impact into three categories: High, Medium, and Low. It represents the difficulty for an attacker to take advantage of a weakness once they have penetrated the target system. A more severe vulnerability would be one with lower complexity, as it would be easier to exploit.

\begin{itemize}
    \item \textbf{High (H)}: Exploiting is highly challenging due to rare configurations or unusual conditions.
    \item \textbf{Medium (M)}: Exploiting requires specific conditions and moderate skills to deceive users.
    \item \textbf{Low (L)}: Exploiting is straightforward, with minimal effort and no advanced skills needed. 
\end{itemize}

\subsubsection{\textbf{Fix generation}}
This task determines if the models can produce accurate and secure patches for vulnerable code. The models only generate corrected code for the explicitly labeled bugs in the code. The aim is to ensure the solution is technically valid, easy to use, and solves the issue.

\subsection{Dataset Exploration}

We utilized two C/C++ based popular software code vulnerability datasets: Big-Vul~\cite{bigvul} and VulRepair~\cite{Chen_2023vulrepair}.

\textbf{\textit{Big-Vul:}} 
This dataset includes 348 open-source GitHub projects and 91 vulnerability types from 2002–2019.
Each entry has 35 features such as severity scores, access complexity, CVEs, and CWE IDs, along with patched code marking defect lines.
Functions containing defect lines are labeled as vulnerable; others are labeled as non-vulnerable.
In total, it has 253,096 non-vulnerable, and 11,823 vulnerable entries, without preprocessing.

\textbf{\textit{VulRepair:}} 
This dataset contains 8,482 pairs of vulnerable C functions and their repairs, derived from CVE-Fixes~\cite{Bhandari_2021_cvefixes} and Big-Vul~\cite{bigvul}, spanning 1,754 open-source projects (1999–2021).
We used the VulRepair version by Huang et al.~\cite{repairempirical2023}, which labels buggy and fixed code with $<BUGS>$–$<BUGE>$ and $<FIXS>$–$<FIXE>$ tokens, significantly improving repair accuracy.



\subsection{Hardware Setup}

We conducted all the experiments using Google Colab Pro, leveraging an NVIDIA A100 GPU with 40 GB of VRAM.

\subsection{Model Loading and Tokenization}

We utilized the `$unsloth$' ~\cite{Unsloth_Documentation} framework to load all our models. 
To load models, we specifically employed the -

\noindent`$FastLanguageModel. from\_pretrained$' method with configurations tailored to our experimental needs, such as setting a maximum sequence length of 4096 tokens and enabling 4-bit quantization.

Since `$unsloth$' does not support DeepseekCoder-V2 model, we loaded it using Hugging Face’s native tokenizer. We used -

\noindent`$AutoTokenizer.from\_pretrained$' and - 

\noindent`$AutoModelForCausalLM.from\_pretrained$'. The AutoTokenizer automatically identifies and instantiates the correct tokenizer class based on the provided model identifier.

For model fine-tuning, we used LoRA PEFT (Low-Rank Adaptation for Parameter-Efficient Fine-Tuning), a technique that fine-tunes large models without changing all parameters. Instead, it adds a few trainable parameters to certain model sections using low-rank matrices. 

\subsection{Prompt Engineering}

We used instruct models for prompt engineering and base models for fine-tuning.
Table~\ref{table:methods_comparison} shows the comparative summary of the techniques for each corresponding task.

\subsubsection{\textbf{Zero-shot}}

The term ``zero-shot learning'' describes the practice of applying previously trained models to a new task without first providing the model with any task-specific training examples. 

\textit{``Determine if the following code snippet has a vulnerability: [Code]. Respond with `Yes' or `No'.''}


\subsubsection{\textbf{Few-Shot}}

The few-shot learning method involves executing an already trained model using a few task-specific instances as input. 
We provide a description and examples to assist the model in understanding the task and the expected response pattern. 
For example:

\textit{``Example 1:  [code]  and the [vulnerability type],  }

\textit{Example 2:  [code]  and the [vulnerability type] } 

\textit{Determine if the following code snippet has a vulnerability: 
[code]. Respond with `Yes' or `No'.}
''


\subsection{Fine-tuning}
Fine-tuning adapts a pre-trained language model to a specific task using large, high-quality labeled datasets, enabling it to learn deeply from domain-specific data. Table \ref{tab:params} describes the key hyperparameters of fine-tuning.

\begin{table*}[ht]
\footnotesize
\centering
\caption{Comparison of Methods for Different Tasks}
\label{table:methods_comparison}
\renewcommand{\arraystretch}{1.2}
\begin{tabular}{|p{2.5cm}|p{4.5cm}|p{4.5cm}|p{4.5cm}|}
\hline
\textbf{Task} & \textbf{Fine-Tuning} & \textbf{Zero-Shot} & \textbf{Few-Shot} \\
\hline
Vulnerability Detection & 
Model is trained on a labeled dataset to identify patterns of vulnerabilities. Fine-tuning helps it to categorize vulnerabilities by recognizing small variations. & 
Without prior training on specific cases, zero-shot identifies widespread vulnerabilities by using existing knowledge. & 
Few-shot refines the model's accuracy with examples showing justification behind vulnerabilities, especially in edge cases. \\
\hline
Severity Prediction & 
Model correlates code properties with severity levels, such as context, bug type, and exploitability. Fine-tuning enhances prediction accuracy. & 
Zero-shot predicts severity by ranking vulnerabilities using knowledge of associations between code and security implications. & 
Few-shot improves the model's understanding of severity by providing examples with labeled severity levels. \\
\hline
Access Complexity Classification & 
Model learns to classify the vulnerability effect on system accessibility. Fine-tuning improves classification accuracy. & 
Zero-shot uses generalized guidelines to classify accessing situations based on existing knowledge. & 
Few-shot guides the model by providing labeled examples, helping it weigh different complexity scores. \\
\hline
Fix Generation & 
Fine-tuning helps the model learn to generate fixes for specific vulnerabilities. & 
Zero-shot directly requests fixes based on general secure coding knowledge. & 
Few-shot provides examples of vulnerable code and corresponding fixes, guiding the model to generate similar solutions. \\
\hline
\end{tabular}
\end{table*}

\begin{table*}[htbp]
\footnotesize
\caption{Model details}
\label{tab:model_details}
\centering
\renewcommand{\arraystretch}{1.2}
{
\begin{tabular}{|l|l|l|l|l|l|p{6cm}|}
\hline
\textbf{Model Family} & \textbf{Model Name} & \textbf{Size} & \textbf{Version} & \textbf{Release Date} & \textbf{Chat‑template} & \textbf{Developer and Description} \\ \hline
\multirow{2}{*}{Llama}
  & Llama           & 8B             & 3.1             & Jul 2024     & llama‑3.1    & Meta—longer and stronger context reasoning \\ \cline{2-7}
  & CodeLlama       & 7B  & 2               & Aug 2023     & llama‑3.1    & Meta—specialized in generating code, and natural language about code \\ \hline
\multirow{2}{*}{Gemma}
  & Gemma           & 7B             & 1, 1.1          & Feb 2024     & chatml       & Google—lightweight for text generation and reasoning \\ \cline{2-7}
  & CodeGemma       & 7B         & 1.1             & Apr 2024     & chatml       & Google—fast at multiple coding tasks and efficient code completion     \\ \hline
\multirow{2}{*}{Qwen}
  & Qwen            & 7B             & 2.5             & Sep 2024     & qwen‑2.5     & Alibaba—great at instruction following, understanding diversity of system prompts \\ \cline{2-7}
  & QwenCoder       & 7B       & 2.5             & Sep 2024     & qwen‑2.5     & Alibaba— exceptional performance in mathematics and  code reasoning and code fixing.   \\ \hline
\multirow{2}{*}{Mistral}
  & Mistral         & 7B             & 3               & Sep 2023     & mistral      & MistralAI—low memory-latency but high performance in both text and code completion tasks \\ \cline{2-7}
  & CodeStral       & 22B            & 1               & May 2024            & mistral      & MistralAI— explicitly efficient for code generation tasks   \\ \hline
\multirow{2}{*}{Deepseek}
  & Deepseek R1     & 8B             & Distill‑Llama   & Jan 2025     & customized   & DeepSeekAI— mimic the knowledge and capabilities from larger model while improving speed and reducing costs.        \\ \cline{2-7}
  & DeepseekCoder   & 6.7B      & 2               & Nov 2023     & customized   & DeepSeekAI— advanced code completion and infilling capability  \\ \hline
\end{tabular}
}
\end{table*}



\subsubsection{Chat-template}
Chat templates define the conversational structure using special tokens~\cite{feng2025cgptuningstructureawaresoftprompt}.
We aligned each model with its expected template; i.e., ChatML uses $<|im\_start|>$/$<|im\_end|>$, while Mistral uses $<s>[INST]{prompt}[/INST]$.
Using `$unsloth$` templates like `$get\_chat\_template$`, we standardized conversation flow and preserved context. Any deviation from the established chat template that the base model was trained on can disrupt the model's learning patterns.
For models such as deepseek, codestral, and gemma, `$AutoTokenizer$' from Transformers automatically adapted the format, ensuring consistent and reliable outputs.




\begin{table}[ht]
\scriptsize
\centering
\caption{Key Parameters for Model Fine‑tuning}
\label{tab:params}
\small
\setlength{\tabcolsep}{2pt}
{
  \begin{tabular}{|p{2.5cm}|p{3cm}|p{2.5cm}|}
    \hline
    \textbf{Parameter} & \textbf{Value} & \textbf{Description} \\ \hline
    LoRA Rank (r)      & 8              & LoRA adaptation layer rank. \\ \hline
    LoRA Alpha         & 16             & Update LoRA scaling factor. \\ \hline
    LoRA Dropout       & 0.1            & LoRA layer dropout rate.. \\ \hline
    Target Modules     & q\_proj, k\_proj, v\_proj, o\_proj, gate\_proj, up\_proj, down\_proj, embed\_tokens, lm\_head & LoRA application modules. \\ \hline
    Max Sequence Length & 4096          & Maximum input token length. \\ \hline
    Batch Size (Per Device) & 2         & Samples per device batch. \\ \hline
    Gradient Accumulation Steps & 4     & Steps per weight update. \\ \hline
    Learning Rate(LR)      & $3\times10^{-4}, 2\times10^{-4}$ & Initial learning rate. \\ \hline
    Warmup Steps       & 5              & LR warmup duration. \\ \hline
    Max Training Steps & 800, 1000      & Total training steps. \\ \hline
    Optimizer          & paged\_adamw\_8bit, adamw\_8bit & Training optimizer. \\ \hline
    Weight Decay       & 0.01           & Regularization term to prevent overfitting. \\ \hline
    LR Scheduler  & Cosine         & LR decay strategy. \\ \hline
  \end{tabular}
}

\end{table}

\subsection{Dataset Preprocessing}
We balanced the Big-Vul dataset using the undersampling technique across the first three tasks to ensure better learning~\cite{chen2023diversevul},~\cite{feng2025cgptuningstructureawaresoftprompt}. However, we left the VulRepair dataset for task 4 (fix generation) unchanged, as it contains no specific classes.

\begin{figure}[!htbp]
    \centering
    \includegraphics[width=\linewidth]{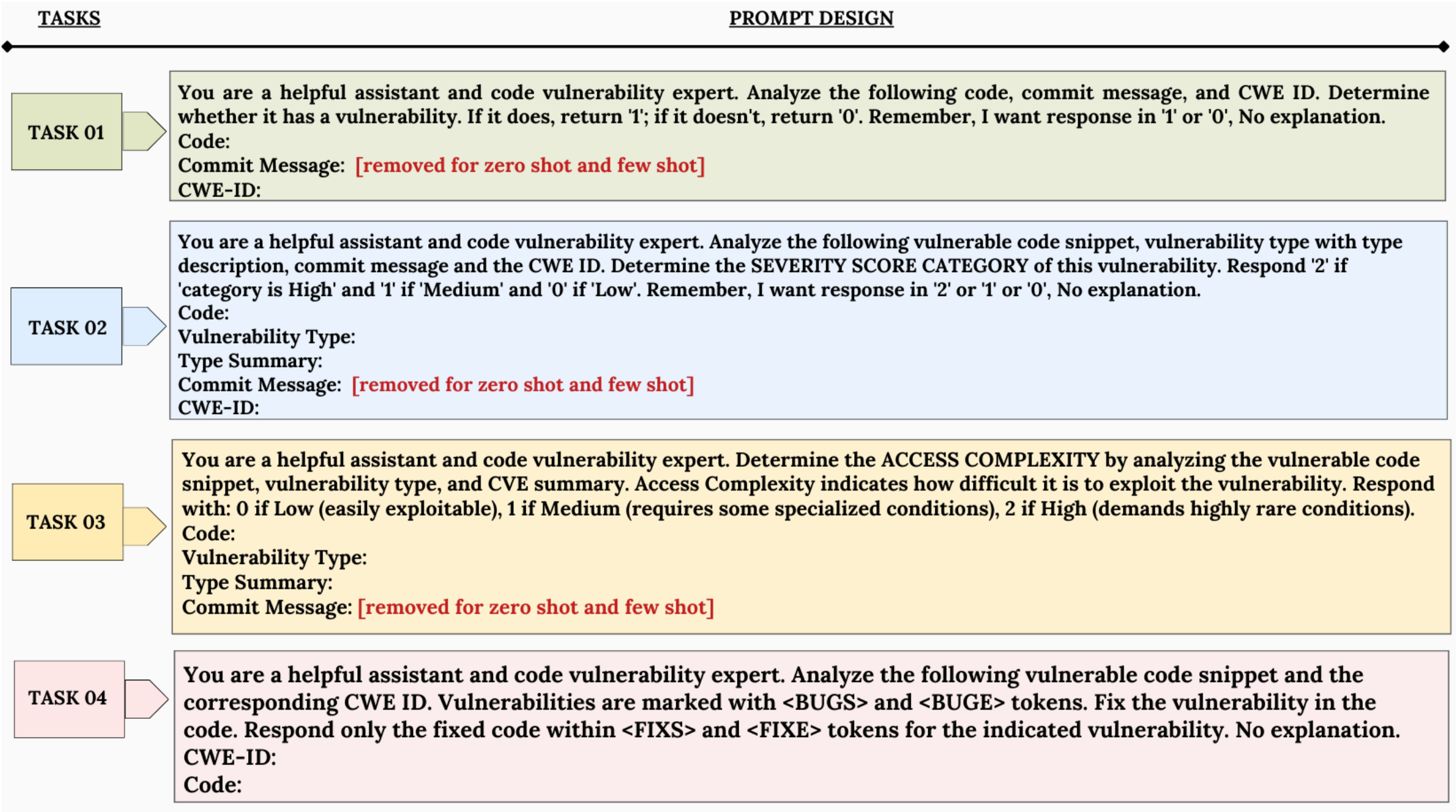}
    \caption{Prompt Design}
    \label{fig:prompt-design}
\end{figure}

We designed concise, resource-efficient prompts and adjusted them (Figure~\ref{fig:prompt-design}) to improve model responses. Prompts were added to the dataset and used in experiments, focusing on vulnerable code for tasks 2 and 3.

For zero- and few-shot setups, each prompt included code, vulnerability type, description, and CWE-ID. Commit messages were removed for clarity and efficiency which also helps reduce the token count and speeds up inference but retained during fine-tuning for added context.


Since most samples in the dataset contain up to 220 lines of code, we applied this limit in zero-shot and few-shot to minimize token usage and processing time. In few-shot, we selected two examples for tasks 1 and 4, and three for tasks 2 and 3. We ensured to add one example from each of the three classes (High, Medium, and Low) that are present in Tasks 2 and 3.

\section{Evaluation} \label{sec:data}

\subsection{Response Extraction}

Extracting desired responses from LLM outputs is challenging and varies across models, tasks, and prompt designs.
Simple tasks like binary classification yield structured outputs, while complex tasks, such as zero-shot severity prediction, often produce verbose explanations, making precise value extraction difficult.
To address this inconsistency, we created and tested a range of regular expressions (regex) to improve the response parsing. The most used regex are:

\lstset{
  basicstyle=\ttfamily\small,
  breaklines=true,
  columns=fullflexible,
  language=Python,
  frame=none,                
  backgroundcolor=\color{white}  
}

\begin{itemize}
    \item {\begin{lstlisting}
match = re.compile(r"(?:assistant\s*(?:answer\s*:\s*)?|answer\s*:\s*)(\d+)", re.IGNORECASE)
    \end{lstlisting}}
    
    \item {\begin{lstlisting}
match = re.search(r"assistant\s*(\d+)", output)
    \end{lstlisting}}
    
    \item {\begin{lstlisting}
match = re.search(r"<think>\s*(\d+)", output)
    \end{lstlisting}}
\end{itemize}
The first regex extracts the response that follows either ``assistant answer:'' or ``answer:'' with flexible spacing and case, and returns the value. The remaining regexes are similar, with slight variations based on different response patterns.

\subsection{Performance Metrics}
For detection, metrics such as precision and recall are crucial, and striking a balance between them is necessary. Overemphasis on precision may result in missing vulnerabilities (poor recall), while excessive focus on recall could cause an excess of false positives (low precision). 

For prediction, weighted F1 scores are vital to ensure fair evaluation of all classes. It gives a balanced view of the performance. A high F1 score is expected, while a low score means that the model is struggling with less common classes, such as missing high-severity vulnerabilities but detecting low-severity ones only. 

For code fix generation, CodeBERTScore~\cite{zhou-etal-2023-codebertscore}, CodeBLEU~\cite{codebleu2020}, ChrF~\cite{chrf}, BLEU‑4~\cite{Bleu4}, and ROUGE‑L~\cite{rouge} are commonly used metrics to evaluate the quality of generated code. Low values in this metric indicate ineffective or insecure fixes ~\cite{enhancingreverseengineering2024}, ~\cite{islam2024enhancingsourcecodesecurity}. Together, these metrics offer a balanced evaluation suite: embedding‑based semantic matching (CodeBERTScore), syntax‑ and semantics‑aware n‑gram overlap (CodeBLEU), fine‐grained character matching (CHRF), strict n‑gram precision (BLEU‑4), and subsequence recall (ROUGE‑L), enabling more nuanced assessment of generated outputs in both code and natural language tasks.

\section{Results} \label{sec:results and findings}

\subsection{\textbf{RQ1: Fine-tuning vs prompt}}

To address this question, we consider solely Task 1, the fundamental vulnerability detection task. Table~\ref{rq1} presents the F1 scores for task 1 across all approaches. We focus on F1 only, as it reflects both precision and recall. According to the results, the fine-tune method is the most effective strategy, outperforming both the zero-shot and few-shot approaches. 

According to our experimental hardware setup and all models featuring the 7B architecture (except Llama3.1-8B and CodeStral-22B), the fine-tune approach generally outperforms both zero-shot and few-shot methods. Table~\ref{rq1} shows that most models exhibit their highest performance scores under fine-tuning. Even though CodeLlama and CodeGemma show competitive scores (76\% and 79\%, respectively) in zero-shot and few-shot conditions, fine-tuning appears to be more consistent and provides robust performance overall. DeepSeek R1 exhibits the highest score of 93\% in fine-tuning for task 1. Nonetheless, Llama and Mistral show optimum results ranging from 70\%-85\% for fine-tuning and 60\%-64\% for prompt-based approaches. Zero-shot and few-shot approaches rely on the pretrained model knowledge without any extra training on the specific task. Our experimental results suggest that few-shot learning performs similarly to zero-shot learning, despite the assumption that it would outperform zero-shot learning.

\begin{table}[ht]
\footnotesize
\centering
\caption{Performance Score (F1) for Task 1}
\label{rq1}
\begin{tabular}{lccc}
\toprule
Model          & Fine-Tune & Zero-Shot & Few-Shot \\
\midrule
Llama3.1-8B          & 0.70   & 0.64   & 0.64   \\
CodeLlama-7B      & 0.75   & \textbf{0.76}  & 0.56   \\
Gemma-7B          & 0.73   & 0.38   & 0.75   \\
CodeGemma-7B      & 0.76   & 0.54   & \textbf{0.79}   \\
Qwen2.5-7B           & 0.67   & 0.58   & 0.38   \\
Qwen2.5 Coder-7B      & 0.74   & 0.32   & 0.30   \\
Mistral-7B        & 0.85   & 0.61   & 0.61   \\
Codestral-22B      & 0.79   & 0.57   & 0.48   \\
DeepSeek R1    & \textbf{0.93}   & 0.70   & 0.63   \\
DeepSeek V2Coder-6.7B  & 0.72   & 0.65   & 0.45   \\
\bottomrule
\end{tabular}
\end{table}

\tcbset{
  answerbox/.style={
    colback=gray!20,      
    colframe=black!50,     
    arc=1pt,              
    left=1mm, right=1mm,  
    top=1mm, bottom=1mm,  
  }
}
\begin{tcolorbox}[answerbox]
\small
\textbf{RQ1:} Fine-tuning emerges as the most effective strategy, despite zero-shot and few-shot methods being more resource-efficient. Few-shot learning performs on par with zero-shot, contrary to expectations.
\end{tcolorbox}

\subsection{\textbf{RQ2: Code-specific models vs General-purpose models}}

\begin{figure}[!htbp]
    \centering
    \includegraphics[width=0.9\linewidth]{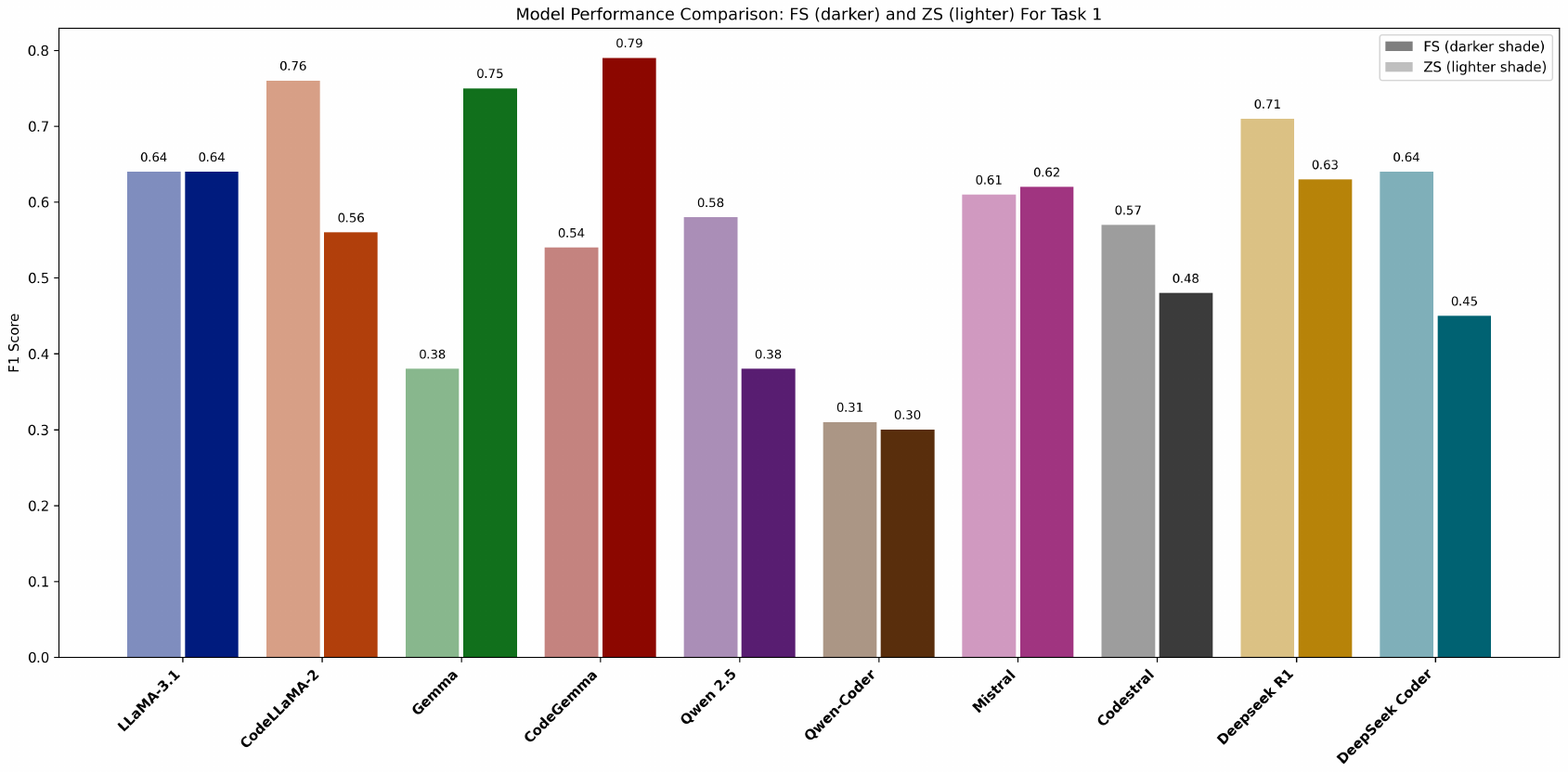}
    \caption{Model Performance Comparison based on Zero-Shot and Few-Shot approach for task 1}
    \label{fig:ZS_FS_1}
\end{figure}

\begin{figure}[!htbp]
    \centering
    \includegraphics[width=0.9\linewidth]{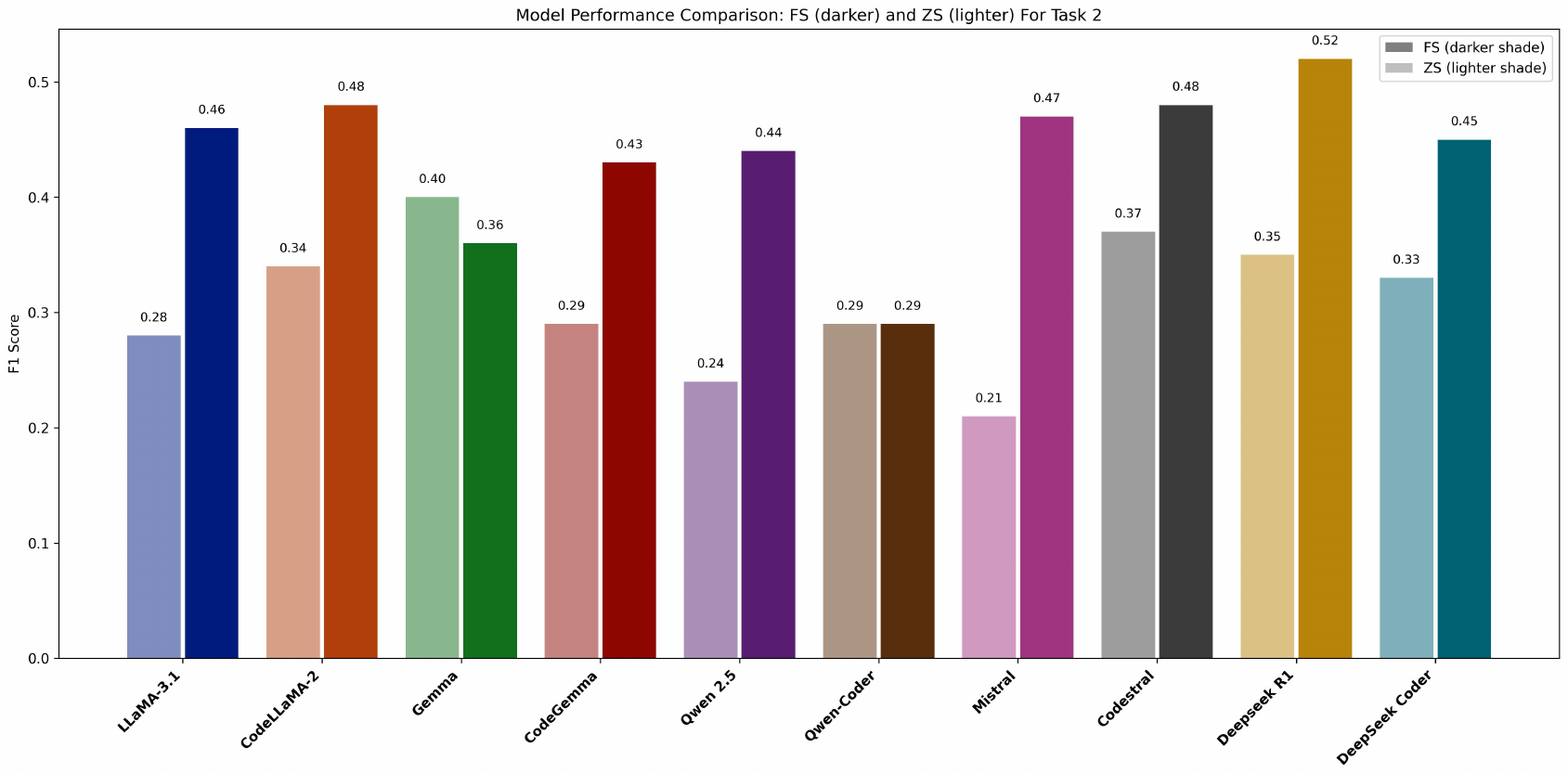}
    \caption{Model Performance Comparison based on Zero-Shot and Few-Shot approach for task 2}
    \label{fig:ZS_FS_2}
\end{figure}

\begin{figure}[!htbp]
    \centering
    \includegraphics[width=0.9\linewidth]{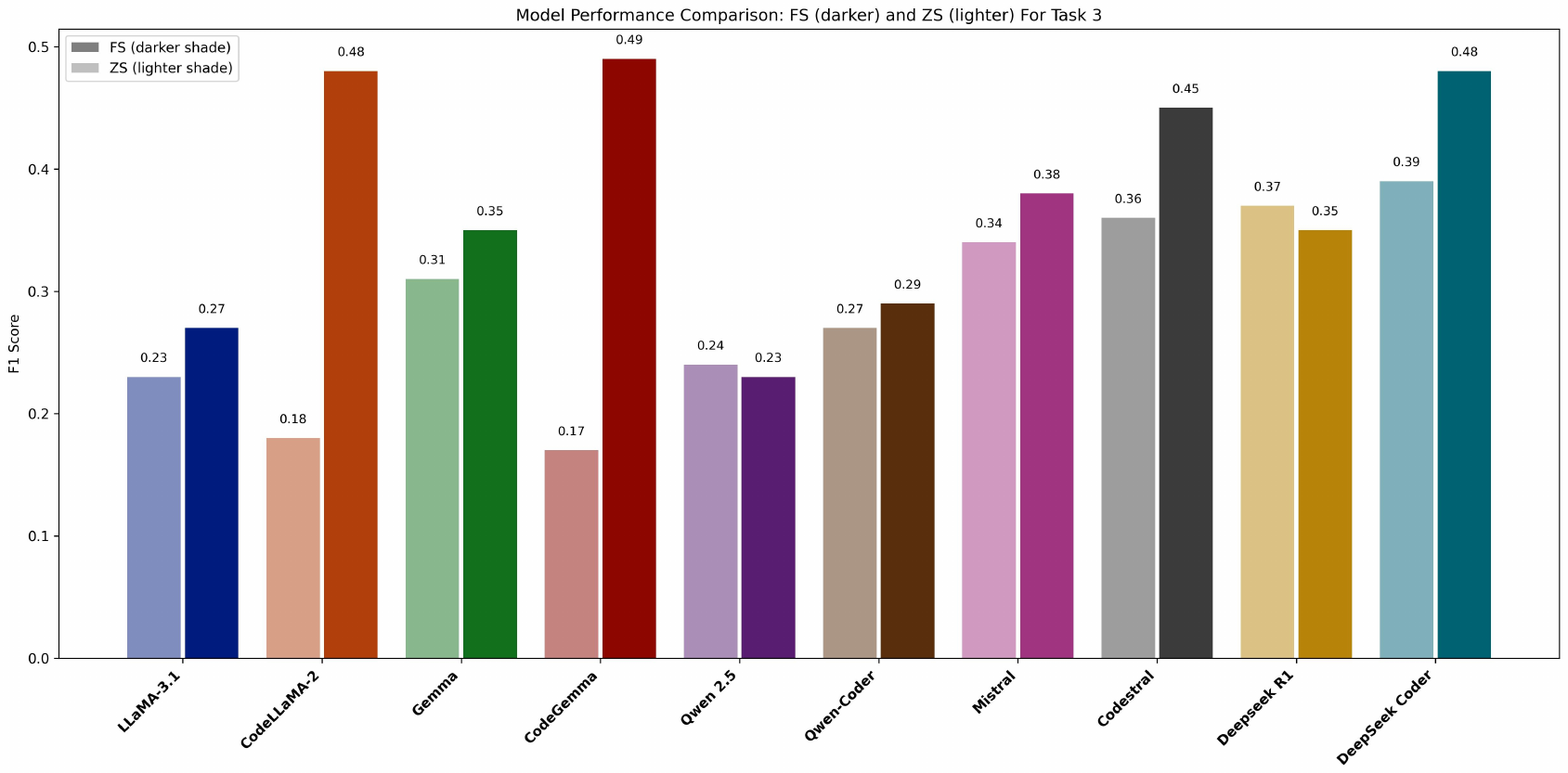}
    \caption{Model Performance Comparison based on Zero-Shot and Few-Shot approach for task 3}
    \label{fig:ZS_FS_3}
\end{figure}

Tables \ref{rq2} and \ref{code_gen_compare} and Figures \ref{fig:ZS_FS_1}, \ref{fig:ZS_FS_2}, and \ref{fig:ZS_FS_3} compare the performance difference in F1 score between code-specific models and general-purpose models across three tasks using fine-tune, zero-shot, and few-shot approaches. In Table~\ref{code_gen_compare}, positive values indicate that code-specific models performed better. According to all the results, there is no clear overall winner model. However, in fine-tuning on Task 1 and Task 2, code-specific models show slightly better performance. But with zero-shot and few-shot methods, general-purpose models often perform equally well or better. Our findings indicate that, across all approaches, three out of five code-specific models exhibit a marginally superior performance than the general models.
Notably, DeepSeek R1, a general-purpose model, consistently outperformed DeepSeek Coder-V2 across all tasks and approaches. This implies that not all code-specific models automatically gain an advantage, especially when pretrained models like DeepSeek R1 are already stronger in general language understanding and adaptation than the V2 version of the code-specific model.

While the zero-shot approach shows balanced results between code-specific and general models on average, the few-shot setting reveals some trends. In few-shot, all the code-specific models underperformed in Task 1 compared to their general-purpose models. However, this is reversed in Tasks 2 and 3, where all code-specific models outperformed. In the few‑shot setting, code‑specific models’ performance fell by 18\% on task 1 but rose to 21\% on task 3. This suggests that while few-shot learning may not be highly effective for more straightforward tasks, it can be more beneficial in complex scenarios. The prior code-specific knowledge plays a larger role, along with the few-shot examples given in this case. This observation also highlights that model performance can vary not only by model type but also by the types of tasks and the task-specific guidance. The findings here, for the most part, suggest that code-specific models do not consistently outperform general-purpose models across all tasks and approaches. The effectiveness depends on the task type and the approach used.

\begin{table}[ht]
\footnotesize
\centering
\caption{Performance Score (F1) for Fine-Tune}
\label{rq2}
\begin{tabular}{lccc}
\toprule
Model          & Task 1 & Task 2 & Task 3 \\
\midrule
Llama3.1-8B          & 0.70   & 0.78   & 0.76   \\
CodeLlama-7B      & 0.75   & 0.79   & \textbf{0.79}   \\
Gemma-7B          & 0.73   & 0.65   & 0.75   \\
CodeGemma-7B      & 0.76   & 0.70   & \textbf{0.79}   \\
Qwen2.5-7B           & 0.67   & 0.76   & 0.76   \\
Qwen2.5 Coder-7B      & 0.74   & 0.80   & 0.75   \\
Mistral-7B        & 0.85   & \textbf{0.84}   & \textbf{0.79}   \\
Codestral-22B      & 0.79   & 0.63   & 0.71   \\
DeepSeek R1    & \textbf{0.93}   & 0.82   & 0.75   \\
DeepSeek V2Coder-6.7B  & 0.72   & 0.74   & 0.74   \\
\bottomrule
\end{tabular}
\end{table}

\definecolor{maroon}{rgb}{0.5,0,0}
\definecolor{darkgreen}{rgb}{0,0.5,0}

\newcommand{\newpos}[1]{\textcolor{maroon}{#1}}
\newcommand{\newneg}[1]{\textcolor{darkgreen}{#1}}

\begin{table}[ht]
\footnotesize
\centering
\caption{F1 score difference for Fine-Tune, Zero-Shot, and Few-Shot approaches across tasks T1, T2, and T3. Positive values mean code-specific models have better performance than general-purpose ones.}
\label{code_gen_compare}
\setlength{\tabcolsep}{10pt}
\renewcommand{\arraystretch}{1.2}
\begin{tabular}{|l|c|c|c|}
\hline
\multicolumn{4}{|c|}{\textbf{Code specific vs General purpose LLM performance(\%)}} \\ \hline
Model Name & T1 & T2 & T3 \\ \hline
\multicolumn{4}{|c|}{\textbf{Fine-Tune}} \\ \hline
CodeLlama-7B      & \newneg{0.05} & \newneg{0.01} & \newneg{0.03} \\
CodeGemma-7B      & \newneg{0.03} & \newneg{0.05} & \newneg{0.04} \\
Qwen2.5 Coder-7B      & \newneg{0.07} & \newneg{0.04} & \newpos{-0.01}  \\
Codestral-22B      & \newpos{-0.06}  & \newpos{-0.21}  & \newpos{-0.08}  \\
DeepSeek V2Coder-6.7B & \newpos{-0.21}  & \newpos{-0.08}  & \newpos{-0.01}  \\ \hline
\multicolumn{4}{|c|}{\textbf{Zero-Shot}} \\ \hline
CodeLlama-7B      & \newneg{0.12} & \newneg{0.06} & \newpos{-0.05} \\
CodeGemma-7B      & \newneg{0.16} & \newpos{-0.11}  & \newpos{-0.14} \\
Qwen2.5 Coder-7B      & \newpos{-0.27}  & \newneg{0.05} & \newneg{0.03} \\
Codestral-22B      & \newpos{-0.04}  & \newneg{0.16} & \newneg{0.02} \\
DeepSeek V2Coder-6.7B & \newpos{-0.07}  & \newpos{-0.02}  & \newneg{0.02} \\ \hline
\multicolumn{4}{|c|}{\textbf{Few-Shot}} \\ \hline
CodeLlama-7B      & \newpos{-0.08}  & \newneg{0.02} & \newneg{0.21} \\
CodeGemma-7B      & \newpos{-0.09}  & \newneg{0.07} & \newneg{0.14} \\
Qwen2.5 Coder-7B      & \newpos{-0.08}  & \newpos{-0.15}  & \newneg{0.06} \\
Codestral-22B      & \newpos{-0.14}  & \newneg{0.01} & \newneg{0.07} \\
DeepSeek V2Coder-6.7B & \newpos{-0.18}  & \newpos{-0.07}  & \newneg{0.13} \\ \hline
\end{tabular}
\end{table}

\begin{tcolorbox}[answerbox]
\small
\textbf{RQ2:} Code-specific models do not consistently exceed general-purpose models in certain tasks and methodologies.  Both the complexity of the task and the approach influence its effectiveness.
\end{tcolorbox}

\subsection{\textbf{RQ3: Superior performing model pair}}

Table~\ref{rq3} presents the average F1 scores of model pairs across tasks and approaches. Due to the difference in performance metrics, we consider excluding code repair task performance. The bold-colored values indicate the highest duo average F1 score for a specific task and approach. 
We can clearly see that Llama and Deepseek duo stand out as the most consistent ones, achieving the highest scores in the majority of cases. Llama3.1 and CodeLlama pair obtain an average of 78\% in fine-tuning tasks and 70\%-60\% in zero-shot and few-shot settings. Additionally, DeepSeek R1 and DeepSeekCoder V2 pair meet 78\%-83\% in fine-tuning and moderate gain 42\%-48\% in prompt-based approaches.
Nevertheless, although Mistral excels independently on average, the duo's ranking has declined due to its ties with Codestral, which frequently underperforms. 

Hence, the Llama and Deepseek models are the ones that generally perform well overall in these kinds of scenarios.

\begin{tcolorbox}[answerbox]
\small
\textbf{RQ3:} Across every task and approach, Llama and DeepSeek models are consistently good performers.
\end{tcolorbox}

\definecolor{darkgreen}{rgb}{0,0.5,0}
\newcommand{\highest}[1]{\textcolor{darkgreen}{#1}}

\begin{table*}[ht]
\footnotesize
\centering
\caption{Average F1 scores for pair of models and tasks under Fine-tune, Zero-Shot, and Few-Shot settings.}
\label{rq3}
\renewcommand{\arraystretch}{1.1}
\begin{tabular}{|c|c|c|c|c|c|c|c|c|c|c|}
\hline
\multirow{3}{*}{Model} & \multicolumn{3}{c|}{Finetune} & \multicolumn{3}{c|}{Zero-Shot} & \multicolumn{3}{c|}{Few-Shot} \\ \cline{2-10}
                       & Task 1    & Task 2    & Task 3    & Task 1    & Task 2    & Task 3   & Task 1    & Task 2    & Task 3 \\ \hline
Llama                  & 0.72   & \highest{\textbf{0.78}}   & \highest{\textbf{0.78}}   & \highest{\textbf{0.70}}   & 0.31   & 0.20   & \highest{\textbf{0.60}}   & 0.47    & 0.37  \\ \hline
Gemma                  & 0.75   & 0.68    & 0.77   & 0.46    & \highest{\textbf{0.35}}    & 0.24   & 0.57    & 0.39    & \highest{\textbf{0.42}} \\ \hline
Qwen                   & 0.70    & \highest{\textbf{0.78}}   & 0.76   & 0.45    & 0.26   & 0.25    & 0.34   & 0.36   & 0.26  \\ \hline
Mistral                & 0.82   & 0.73    & 0.75    & 0.59   & 0.29   & 0.35   & 0.55   & \highest{\textbf{0.48}}   & 0.41 \\ \hline
Deepseek               & \highest{\textbf{0.83}}   & \highest{\textbf{0.78}}    & 0.74   & 0.68     & 0.34      & \highest{\textbf{0.42}}    & 0.54   & \highest{\textbf{0.48}}     & 0.41   \\ \hline
\end{tabular}
\end{table*}

\subsection{\textbf{RQ4: Evaluation of code repair similarity metrics}}

For RQ4, we evaluate our models with five mostly used metrics on the C program-based VulRepair dataset. Table~\ref{fixgen} presents that fine-tuning yields better gains across all the metrics. The red color indicates a low score, while green indicates a high score for each case. However, zero-shot and few-shot learning lag behind mostly due to incomplete context understanding. It is clear from the table that DeepSeek R1 stands out as the top model in every case, with Mistral also exhibiting good performance. Additionally, in a few-shot setting, Qwen2.5-Coder achieves impressive scores. Though zero-shot performance is the weakest, it is the best to test a model's general understanding.
Additionally, examining the metrics reveals that CodeBERTScore and Rouge-L consistently yield higher scores across all cases. 

However, BLEU‑4 and CodeBLEU scores stay relatively low even when other metrics are strong, which raises questions about how well BLEU-based metrics capture the actual improvements or meaning in code generation tasks. 
Although code-specific, CodeBLEU and CodeBERTScore results reveal a measurable difference, which raises concerns about their reliability in assessing an LLM's overall capability.
The similarity metrics only measure how much a generated fix matches the reference, not whether it truly works. Models may still be able to fix bugs even if their suggestions do not precisely match the truth. However, depending on a single ground truth is often inadequate to determine whether the LLM is producing accurate results. Furthermore, modifying code in C generally involves rearranging several lines of code, which can be done in various ways. This flexibility complicates code matching and increases the risk of bugs. Python’s conciseness significantly simplifies comparison and reduces evaluation complexity.

Considering all approaches, we observe that the highest CodeBERTScore, Rouge-L, ChrF, and BLEU‑4 gains are 89\%, 60\%, 42\%, 33\% by DeepSeek R1. Additionally, the highest CodeBLEU score is 37\% by Llama3.1.
Overall, these findings suggest that task-specific fine-tuning plays a significant role in improving code-generation quality. Moreover, considering that zero-shot results remain relatively low, the few-shot setup might not be able to capture a significant amount of the fine-tuned strength. Combining hands‑on testing with selective expert assessment can provide a clear picture of whether the LLM's fix generation is suitable.

\begin{tcolorbox}[answerbox]
\small
\textbf{RQ4:} Metrics often used for LLM-generated vulnerability repairs prioritize syntactic similarity over functional correctness, which necessitates an evaluation strategy towards semantic accuracy to effectively generalize across different programming languages. Task-specific fine-tuning significantly enhances code repair quality; however, few-shot and zero-shot setups struggle to capture context and yield relatively modest results. \end{tcolorbox}

\definecolor{maroon}{rgb}{0.5,0,0}
\definecolor{darkgreen}{rgb}{0,0.5,0}

\newcommand{\low}[1]{\textcolor{maroon}{#1}}
\newcommand{\high}[1]{\textcolor{darkgreen}{#1}}

\begin{table}[ht]
\footnotesize
\centering
\caption{Fix Generation Performance}
\label{fixgen}
\setlength{\tabcolsep}{2pt}
\renewcommand{\arraystretch}{1.2}
\begin{tabular}{|l|c|c|c|c|c|}
\hline
Model Name & CodeBert & Bleu-4 & CodeBleu & Rouge-L & ChrF \\ \hline

\multicolumn{6}{|c|}{\textbf{Fine-Tune}} \\ \hline
Llama3.1-8B      & 0.85 & 0.15 & \low{0.19} & 0.53 & 0.35 \\
CodeLlama-7B  & 0.85 & 0.16 & 0.20 & 0.50 & 0.29 \\
Gemma-7B      & \low{0.82} & 0.17 & 0.23 & 0.49 & 0.30 \\
CodeGemma-7B   & 0.84 & \low{0.13} & 0.20 & \low{0.48} & \low{0.28} \\
Qwen2.5-7B   & 0.85 & 0.19 & 0.21 & 0.52 & 0.34 \\
Qwen2.5 Coder-7B  & 0.84 & 0.19 & 0.23 & \low{0.48} & 0.35 \\
Mistral-7B    & 0.87 & 0.14 & 0.22 & 0.57 & 0.36 \\
Codestral-22B & 0.85 & \low{0.13} & 0.20 & 0.55 & 0.33 \\
DeepSeek R1      & \high{0.89} & \high{0.33} & \high{0.34} & \high{0.60} & \high{0.42} \\
DeepSeek V2Coder-6.7B    & 0.86 & 0.17 & 0.24 & 0.53 & 0.35 \\ \hline

\multicolumn{6}{|c|}{\textbf{Zero-Shot}} \\ \hline
Llama3.1-8B      & 0.72 & 0.18 & 0.11 & 0.26 & 0.25 \\
CodeLlama-7B  & 0.69 & \high{0.19} & 0.12 & 0.24 & 0.22\\
Gemma-7B      & \low{0.64} & 0.12 & 0.11 & \low{0.10} & \low{0.12} \\
CodeGemma-7B   & 0.68 & \low{0.11} & 0.14 & 0.19 & 0.26  \\
Qwen2.5-7B   & 0.74 & 0.13 & 0.10 & 0.37 & 0.22 \\
Qwen2.5 Coder-7B  & 0.73 & 0.15 & \low{0.09} & 0.36 & 0.21 \\
Mistral-7B    & \low{0.64} & \high{0.19} & \high{0.15} & 0.16 & 0.19 \\
Codestral-22B & 0.66 & 0.18 & 0.13 & 0.26 & 0.25 \\
DeepSeek R1    & \high{0.75} & 0.16 & 0.12 & \high{0.47} & \high{0.28} \\
DeepSeek V2Coder-6.7B  & 0.71 & 0.15 & 0.10 & 0.42 & 0.24\\ \hline

\multicolumn{6}{|c|}{\textbf{Few-Shot}} \\ \hline
Llama3.1-8B      & 0.65 & 0.13 & \high{0.37} & 0.27 & \low{0.15} \\
CodeLlama-7B  & 0.58 & \low{0.10} & 0.32 & \low{0.21} & 0.18 \\
Gemma-7B      & \low{0.56} & 0.12 & 0.13 & \low{0.21} & 0.17 \\
CodeGemma-7B   & 0.59 & 0.14 & 0.16 & 0.24 & 0.16 \\
Qwen2.5-7B   & 0.78 & \high{0.20} & 0.21 & \high{0.52} & \high{0.34} \\
Qwen2.5 Coder-7B  & 0.79 & 0.14 & 0.15 & 0.41 & 0.25 \\
Mistral-7B    & 0.73 & 0.18 & 0.13 & 0.25 & 0.26 \\
Codestral-22B & 0.72 & 0.15 & \low{0.11} & 0.27 & 0.22 \\
DeepSeek R1   & \high{0.80} & 0.18 & 0.19 & 0.34 & 0.23\\
DeepSeek V2Coder-6.7B   & 0.75 & 0.16 & 0.17 & 0.31 & 0.19 \\ \hline

\end{tabular}
\end{table}

\section{Threats to Validity} \label{sec:limitations}
\begin{itemize}
\item Occasional ``None'' outputs appeared in zero- and few-shot runs. Models like Qwen and Mistral handled formats better than others, such as Gemma, though some code-general pairs showed inconsistent behavior.
\item Benchmarks such as Big-Vul contain mislabeled samples~\cite{croft2023dataquality}, which can misrepresent model accuracy, especially in zero-shot cases.
\item Instruction-tuned LLMs often generated verbose outputs, requiring higher `$max\_new\_tokens$`, longer inference times, and extra post-processing.
\item Experiments were limited to C-language repairs and selected classification tasks; generalization to other languages or domains remains uncertain. The findings establish a baseline for LLM-based software vulnerability workflow with future potential for memory-safe languages.
\end{itemize}

\section{Discussions \& Conclusion}
We evaluate the usefulness of LLMs as integrated assistants throughout the SVM pipelines by analyzing them across four key sections: identifying vulnerabilities, prioritizing risks based on severity, assessing access complexity, and finally, resolving identified threats. Our results highlight the value of task-specific fine-tuning for maximizing LLM performance in vulnerability-related tasks. Fine-tuned models consistently outperformed zero- and few-shot setups, showing that pretrained representations benefit from targeted adaptation, especially for repair generation.

Comparing code-specific and general models revealed that general models often outperform specialized ones in zero- and few-shot settings, suggesting that broad language knowledge can offset limited domain pretraining. Model choice should depend on the deployment context, and ensembling may improve reliability.

Although fine-tuning large models requires significant resources, our experiments achieved strong results under limited compute. The study also exposed limitations in reference-based evaluation metrics, such as the fact that token overlap does not always reflect true functional accuracy. Future work should incorporate execution-based testing to better assess security effectiveness.

Overall, we measured open-source LLMs across four vulnerability tasks using fine-tuned, zero-shot, and few-shot methods. Fine-tuning consistently yielded the best performance, while code-specific pretraining offered no clear advantage. Future directions include multi-language expansion, and ensemble strategies for robust vulnerability management.

\bibliographystyle{abbrv}
\bibliography{references} 
\end{document}